\documentclass[a4paper,11pt]{article}
\usepackage{jcappub} % for details on the use of the package, please
\usepackage[T1]{fontenc} % if needed

\usepackage{ulem}
\input{colordvi.tex}
\usepackage{aas_macros}

\title{\boldmath Primordial magnetic fields from self-ordering scalar fields}

\author[a,1]{Kouichirou Horiguchi,\note{Corresponding author.}}
\author[a,b]{Kiyotomo Ichiki,}
\author[c]{Toyokazu Sekiguchi}
\author[a,b,d]{and Naoshi Sugiyama}

\affiliation[a]{Department of Physics and Astrophysics, Nagoya University,\\ Aichi 464-8602, Japan}
\affiliation[b]{Kobayashi-Maskawa Institute for the Origin of Particles and the Universe, \\Nagoya University, \\Nagoya 464-8602, Japan}
\affiliation[c]{Helsinki Institute of Physics, University of Helsinki, \\PO Box 64, FIN-00014, Finland}
\affiliation[d]{Kavli Institute for the Physics and Mathematics of the Universe (Kavli IPMU), The University of Tokyo, \\Chiba 277-8582, Japan}
\emailAdd{horiguchi.kouichirou@h.mbox.nagoya-u.ac.jp}
\emailAdd{ichiki@a.phys.nagoya-u.ac.jp}
\emailAdd{toyokazu.sekiguchi@helsinki.fi}
\emailAdd{naoshi@nagoya-u.jp}

\abstract{ A symmetry-breaking phase transition in the early universe
could have led to the formation of cosmic defects. Because these defects
dynamically excite not only scalar and tensor type cosmological
perturbations but also vector type ones, they may serve as a source of
primordial magnetic fields.  In this study, we calculate the time evolution and
the spectrum of magnetic fields that are generated by a type of cosmic
defects, called global textures, using the non-linear sigma (NLSM)
model. Based on the standard cosmological perturbation theory, we show,
both analytically and numerically, that a vector-mode relative velocity
between photon and baryon fluids is induced by textures, which 
inevitably leads to the generation of magnetic fields over a wide range of
scales.  We find that the amplitude of the magnetic fields is given by
$B\sim{10^{-9}}{((1+z)/10^3)^{-2.5}}\left({v}/{m_{\rm
pl}}\right)^2\left({k}/{\rm Mpc^{-1}}\right)^{3.5}/{\sqrt{N}}$ Gauss in
the radiation dominated era for $k\lesssim 1$ Mpc$^{-1}$, with $v$ being
the vacuum expectation value of the O(N) symmetric scalar fields.  By extrapolating our
numerical result toward smaller scales, we expect that $B\sim
{10^{-17}}\left((1+z)/1000\right)^{-1/2}\left({v}/{m_{\rm
pl}}\right)^2\left({k}/{\rm Mpc^{-1}}\right)^{1/2}/{\sqrt{N}}$ Gauss on
scales of $k\gtrsim 1$ Mpc$^{-1}$ at redshift $z\gtrsim 1100$. This
might be a seed of the magnetic fields observed on large scales today.}

\begin{document}
\maketitle
\flushbottom

\section{Introduction}\label{sec:intro} %introduction

%inflation$B$OO@J8$N<g;]$H4X78$J$$$N$G%P%C%5%j(Bcut$B$7$^$7$g$&(B
% Inflation paradigm has been introduced in the standard cosmological
% model to solve problems inherent in the Big-Bang universe, such as
% flatness problem, horizon problem, and others
% \cite{1981MNRAS.195..467S}\cite{1981PhRvD..23..347G}.  In the slow-roll
% inflation scenario, a scalar field called inflaton goes down its
% potential slowly and the accelerated expansion driven by the potential
% energy produces scale-invariant gravitational waves (GWs). Therefore the
% detection of (nearly) scale-invariant gravitational waves has been
% considered as a smoking-gun of the slow-roll inflation in the early
% universe.

The standard big-bang cosmological model has been established by a
variety of astronomical observations, such as the dimming of distant
supernovae, the cosmic microwave background (CMB), the large-scale
structure of the universe, and so on. In this big-bang paradigm, it is
naturally expected that the universe experienced a number of phase
transitions at the early stages of cosmic history, because of cooling
from a very hot initial state caused by its adiabatic
expansion.  As a result of these phase transitions, topological
defects may have been formed and may have acted as a source of large
scale structure in the universe.  Among others, topological defects
resulting from
a phase transition that breaks a global $O(N)$ symmetry are called
$O(N)$ global defects. For example, $O(N)$ symmetric scalar fields can
yield domain walls $(N=1)$, monopoles $(N=3)$, and textures $(N\ge 4)$
when the scalar fields break the global symmetry.

There exist many studies regarding the phenomenological aspects of
topological defects, which include the generation of CMB temperature and
polarization anisotropies \cite{Durrer:1993db,Durrer:1995ni,Durrer:1998rw,Pen:1997ae}, gravitational waves (GWs)\cite{2009JCAP...10..005F,2010PhRvD..82h3518D,JonesSmith:2007ne,2012JCAP...11..006G,2013PhRvL.110j1302F},
and cosmic rays \cite{PhysRevD.36.1007,1994PhRvD..50.3660G}, among others (for a review of
structure formation with 
topological defects, see \cite{2002PhR...364....1D}).  As a rule of
thumb, the amplitude of the fluctuations induced by topological defects,
such as CMB anisotropies, is of the order $\Delta T/T\sim 4\pi Gv^2$,
where $v$ is the vacuum expectation value (VEV) of the scalar fields,
$G$ is the Newton constant, and $T$ and $\Delta T$ are the CMB
temperature and its fluctuation, respectively. Therefore, the recent CMB measurement by the
Planck satellite has placed limits on the energy scale of the topological
defects \cite{Ade:2013xla}. The actual limits depend on the detailed
models, for instance, $Gv^2\leq4.2\times10^{-7}$
\cite{Urrestilla:2011gr} for cosmic strings, $Gv^2\leq3.2\times10^{-7}$
for Abelian--Higgs cosmic strings, $Gv^2\leq1.5\times 10^{-7}$ for
Nambu--Goto strings, and $Gv^2\leq 1.1\times10^{-6}$ for semi-local
strings and global textures \cite{Ade:2013xla}.

%COSMIC DEFECTS
In this paper, we pay particular attention to global textures with 
$N\gg 4$,  
which can be well-approximated by self-ordering scalar
fields that follow the non-linear sigma model (NLSM).
The NLSM can describe the evolution of global $O(N)$ symmetric scalar
fields with an accuracy up to corrections on the order of $1/N$.
The NLSM has attracted much attention since the discovery of CMB B-mode
polarizations by the BICEP2 experiment \cite{Ade:2014xna} because 
textures following the NLSM can be a source of the scale-invariant GWs
\cite{Krauss:1991qu}\cite{JonesSmith:2007ne}\cite{2009JCAP...10..005F},
just as inflation in the early universe can produce the scale-invariant
GWs. To observationally distinguish between GWs originating from inflation and
from textures,
one should consider observables that reflect the time evolution of
the GWs. The GWs from inflation are 
frozen on super-horizon scales at first, and decay with oscillations
after the horizon crossing. The GWs from textures, on the other hand,
are generated inside the horizon, and decay 
with oscillations after the scalar fields that source the GWs decay
away as the 
universe expands.  In \cite{Fenu:2013tea}, the authors calculate the CMB
temperature and polarization anisotropies in the NLSM and find
that the shapes of the correlation functions of the CMB anisotropies in the
NLSM are different from the corresponding ones from inflation. 
Therefore, detailed observations of  CMB anisotropies can distinguish
between GWs from the two different origins. In fact, the recent B-mode 
measurement by BICEP2 places an upper bound on the VEV in the 
NLSM at $v\lesssim 9\times10^{-4} G^{-1/2}$, and the GWs from
the NLSM are shown to be slightly disfavored by the data compared with those from inflation 
\cite{Durrer:2014raa}\cite{Lizarraga:2014xza}, while contamination
by dust in the BICEP2 data has to be re-analyzed with PLANCK data.

In this paper, we investigate yet another route to probe the NLSM:
the 
generation of magnetic fields. Because of the non-linear nature of the
NLSM, scalar fields following the NLSM inevitably induce vector-mode
perturbations as well as scalar and tensor-mode ones.  The relative
vector-mode velocity between photon and baryon fluids induces 
rotation in electric fields, leading to the generation of magnetic fields
\cite{PhysRevD.85.043009}. Recent discoveries of large-scale magnetic
fields in void regions
\cite{2010Sci...328...73N}\cite{2013ApJ...762...15P} as well as 
magnetic fields at  high redshifts \cite{Bernet:2008qp} make the
investigation more interesting because such magnetic fields may be of
primordial origin in the early universe. Therefore, one of the aims of this paper
is to derive the spectrum of magnetic fields generated by 
vector perturbations in the NLSM within the observational limits of
CMB anisotropies.

This paper is organized as follows. In the next section, we review the
NLSM, in which $N$-component scalar fields act as a source of
cosmological perturbations. In section 3, we derive the power spectrum
of magnetic fields in the NLSM, both numerically and analytically, using
the tight coupling approximation. We will see that in order to obtain a
reliable 
result, we should expand the equations up to the third order in the tight
coupling approximation between photon and baryon fluids. We discuss the result and give an analytic
interpretation of the spectrum of the magnetic fields in section 4, followed by our conclusion
in section 5. Throughout this paper, we fix the cosmological parameters
to $h=0.7,\ \Omega_bh^2=0.0226,\ \Omega_{\rm c}h^2=0.112$, and
$N_\nu=3.046$, where $H_0=100h\ {\rm km/s/Mpc}$ is the Hubble constant,
$\Omega_b$ and $\Omega_{\rm c}$ are the density parameters of baryonic and
cold dark matter, respectively, and $N_\nu$ is the effective number of
massless neutrinos.  Those parameter values are consistent with the Planck
results, and correspond to the $\Lambda$CDM model \cite{Ade:2013zuv}.

%There are many observational evidences which indicate  large scale
%magnetic fields \cite{L.M}. Recently, $10^{-16}\sim10^{-20}\ {\rm G}$ is
%found to be the lower bounds of the intergalactic magnetic fields
%\cite{A.Neronov}\cite{K.L}.But its origin has not been  revealed yet. In
%a model, large scale magnetic fields come from seed fields in the early
%universe. One of them say that relative velocity between photon and
%baryon can produce seed field \cite{Kiyotomo}. NLSM can produce
%relative velocity between photon and baryon in the early universe, so it
%is able to produce seed fields.  In this paper, we consider Primordial
%Magnetic Field (PMF) from self-ordering scalar field which follows NLSM.  

\section{Non-linear sigma model (NLSM)}%NLSM
\label{sec:nlsm}%\ref{sec:nlsm} 

We consider a model of $N$ scalar fields with a global $O(N)$ symmetry, which undergoes symmetry breaking in the early universe.
The Lagrangian of the model is given by
\begin{equation}
\label{eq:nlsm_L}%\eqref{eq:nlsm_L}
{\cal L}=-\frac{1}{2}\nabla_{\mu} \Phi^t %^\dag
\nabla^{\mu}\Phi-\frac{\lambda}{4}( \Phi^t %^\dag
\Phi-v^2)^2+{\cal L}_T.
\end{equation}
Here $\Phi %^\dag
=(\phi_1,\phi_2,...,\phi_N)$
is an array of $N$ real scalar fields,  $v$ is the
vacuum expectation value (VEV) of the scalar fields after symmetry breaking, $\lambda$ is
the self-coupling %interaction 
constant, and ${\cal L}_T\sim T^2 \Phi^t \Phi$ is the thermal term of the Lagrangian.
%\KI{
Deep %At first, 
in the %early 
radiation-dominated universe, the thermal term is dominant in 
the Lagrangian of the scalar fields. As the temperature of the universe goes down, 
this thermal term gets smaller. When the term 
becomes negligible 
a spontaneous symmetry breaking 
occurs.
At energy scales well below $v$, 
the field values are confined on the
$N-1$ dimensional sphere 
in the $N$ dimensional field
space so that
$\Phi^t 
(\vec{x},\eta)\Phi(\vec{x},\eta)=v^2$. 
Under this condition, in the large-$N$ limit, the equation of motion
for the scalar fields can be derived as
\begin{equation}
\label{eq:nlsm_EoM}%\eqref{eq:nlsm_EoM}
\nabla^{\mu}\nabla_{\mu}\beta_a+\left(\nabla_{\mu}{\beta_b}\right)\cdot\left(\nabla^{\mu}{\beta_b}\right)\beta_a=0,
\end{equation}
where the indices $a$ and $b$ run over $1, \dots, N$ and summations with respect to deprecated indices are implicit.
This is the equation of motion for
the scalar fields in the NLSM. 
If 
$N \leq 3$, 
topological defects such as domain walls, cosmic strings and
monopoles are produced, where the scalar 
fields can restore 
the $O(N)$ symmetry and possess the 
higher energy density. 

In this paper we consider cases 
with $N\geq4$, where non-topological defects, or textures, form. 
Let us consider the flat Friedmann-Lema\^itre-Robertson-Walker universe with the metric
\begin{equation}
ds^2=g_{\mu\nu}dx^{\mu}dx^{\nu}=a^2(\eta)(-d\eta^2+d\vec{x}^2)~, 
\end{equation}
where
$a(\eta)$ is the cosmic scale factor and $\eta$ is the conformal
time. 
In the large-$N$ limit,
making an ansatz that $\left<
\left(\nabla_{\mu}{\beta_a}\right)\cdot\left(\nabla^{\mu}{\beta_a}\right)\right>=T_0a^{-2}\eta^{-2}$, 
we can obtain the analytic solution as \cite{JonesSmith:2007ne} \cite{2009JCAP...10..005F}
\begin{equation}
\label{eq:nlsm_sol}%\eqref{eq:nlsm_sol}
\beta_a(\vec{k},\eta)=
\sqrt{A_\nu}\left(\frac{\eta}{\eta_*}\right)^{3/2}\frac{J_\nu(k\eta)}{(k\eta)^\nu}\beta_a(\vec{k},\eta_*)\equiv
f(k,\eta,\eta_*)\beta_a(\vec{k},\eta_*),
\end{equation}
where $\nu={\rm d \ln a}/{\rm d \ln \eta}+1$,
$A_\nu=4\Gamma(2\nu-1/2)\Gamma(\nu-1/2)/3\Gamma(\nu-1)$ and 
$T_0= 3\nu-9/4$. In eq.
\eqref{eq:nlsm_sol}, $\eta_*$ is the time 
of phase transition and $\beta_a(\vec{k},\eta_*)$ is the initial value of scalar
fields, whose two-point correlation function 
can be given as
 \begin{eqnarray}
 \label{eq:nlsm_init}%\eqref{eq:nlsm_init}
 \left<
 \beta_a(\vec{k}_1,\eta_*)\beta_b^*(\vec{k}_2,\eta_*)
 \right>
 =%\begin{cases}
 \frac{6\pi^2\eta_*^3}{N}\delta_{ab}\ (2\pi)^3\delta(\vec{k}_1-\vec{k}_2).
%& ({\rm for~} k\eta_*\lesssim 1) \\
%0 & ({\rm for~}k\eta_*\gtrsim 1) 
%\end{cases}.
 \end{eqnarray}
The above relation is only valid for $k\ll1/\eta_*$, which follows, in the
large-$N$ limit, from the fact that the scalar fields take their VEV
independently in each horizon at $\eta_*$.\footnote{ On small scales
$k\gtrsim 1/\eta_*$, initial scalar fields become homogenous and
correlation function of eq. \eqref{eq:nlsm_init} should vanish. We
however note that our argument does not depend on the correlation
function on these scales.}  We also note that $\beta_a(\vec k,\eta_*)$
is Gaussian at these scales.

We denote a correlation function of $\beta_a$ as a product of the
transfer function and the initial amplitude as
\begin{equation}
\label{eq:beta2}
 \left<\beta_a(\vec{k}_1,\eta)\beta_b^*(\vec{k}_2,\eta)\right>= F(k_1,k_2,\eta,\eta_*)\left<\beta_a(\vec{k}_1,\eta_*)\beta_b^*(\vec{k}_2,\eta_*)\right>~,
\end{equation}
where
 $F(k_1,k_2,\eta,\eta_*)\equiv %=
 f(k_1,\eta,\eta_*)f(k_2,\eta,\eta_*)$.
From eq.\eqref{eq:nlsm_sol} and eq.\eqref{eq:nlsm_init}, we can
 see that the dependence of eq.\eqref{eq:beta2} on $\eta_*$ is canceled out. Therefore we omit 
 $\eta_*$ 
 from the equations hereafter and write as
 $F(k_1,k_2,\eta,\eta_*)=F(k_1,k_2,\eta)$ and
 $\beta_a(\vec{k},\eta_*)=\beta_a(\vec{k})$. 
Finally the energy-momentum tensor of the scalar fields is given by
\begin{equation}
\label{eq:nlsm_EMT}%\eqref{eq:nlsm_EMT}
T_{\mu\nu}^{\phi}=v^2\left[\partial_\mu \beta_a\partial_\nu \beta_a-\frac{1}{2}g_{\mu\nu}\partial_\lambda \beta_a\partial^\lambda \beta_a\right].
\end{equation}

\section{Magnetic fields}%magnetic field
\label{sec:magnetic}%{sec:magnetic}
In this section, we investigate generation of seeds of large
scale magnetic fields from the self-ordering scalar fields which 
follow the NLSM. These scalar fields can induce cosmological vector-mode perturbations 
and eventually %hence 
produce magnetic fields.

\subsection{Vector mode perturbations and their evolution equations }
\label{subsec:nlsm_vec}%subsec:nlsm_vec
We begin by reviewing the basic linear perturbation theory and
define the vector mode. 
Let us consider the perturbed metric around the flat FRW one in the synchronous gauge as
\begin{equation}
ds^2=g_{\mu\nu}dx^{\mu}dx^{\nu}=a^2(\eta)(-d\eta^2+(\delta_{ij}+h_{ij})dx^i\ dx^j),
\end{equation}
where 
$h_{ij}$ is  
the metric perturbation. In Fourier space, 
the vector part of $h_{ij}$ can be expressed as
\begin{equation}
\label{eq:h_decomp1}%\eqref{eq:h_decomp1}
h_{ij}=\frac{i\hat{k}_ih^V_j+i\hat{k}_jh_i^V}{\sqrt{2}}.
\end{equation}
Here $h_i^V$ is a divergenceless vector and
can be rewritten using the vector basis
$e_i^{(\pm)}(\hat{k})$ as %, 
\begin{equation}
\label{eq:h_decomp2}%\eqref{eq:h_decomp2}
h_i^V=\sum_{\lambda=\pm}\lambda h^{(\lambda)}_Ve_i^{(\lambda)}(\hat{k}).
\end{equation}
Combining eq. \eqref{eq:h_decomp1} and eq. \eqref{eq:h_decomp2}, we can
denote $h_{ij}$ directly as
\begin{equation}
h_{ij}=\sum_{\lambda=\pm}h_V^{(\lambda)}{\cal O}^{(\lambda)}_{ij},\quad {\cal O}^{(\lambda)}_{ij}=\frac{i\lambda}{\sqrt{2}}(\ \hat{k}_ie_j^{(\lambda)}(\hat{k})+\hat{k}_je_i^{(\lambda)}(\hat{k})\ ),
\end{equation}
where ${\cal O}^{(\lambda)}_{ij}$ is the vector projection
tensor. Using this projection tensor, we can derive the vector-mode
perturbation equation for $\sigma=\dot{h}_V/k$ as
\begin{equation}
\label{eq:EoMS}%\eqref{eq:EoMS}
\dot{\sigma}^{(\lambda)}+2{\cal H}\sigma^{(\lambda)}=8\pi G a^2 \Pi^{(\lambda)}/k,
\end{equation}
where $\Pi^{(\lambda)}=T^{(\rm{tot})}_{ij}{\cal O}_{ij}^{(\lambda)}$ is the total anisotropic stress in the
vector mode,${\cal H}=\dot{a}/a$ is conformal hubble, and a dot denotes a
conformal time derivative. Hereafter we shall omit the superscript
${(\lambda)}$ for the purpose of presentation. 
The total energy-momentum tensor consists of two parts: one is
from the ordinary matter and radiation and the other is from the scalar fields
given by eq. \eqref{eq:nlsm_EMT}.
 Anisotropic stress of the scalar fields
 $\Pi^\phi=T^{\phi}_{ij}{\cal O}_{ij}$ can be calculated as
\begin{equation}
\label{eq:nlsm_AS}%\eqref{eq:nlsm_AS}
a^2\Pi^{\phi}(\vec{k},\eta)=\frac{v^2}{2}\int \frac{{\rm d}^3p}{(2\pi)^3}\frac{{\rm d}^3q}{(2\pi)^3}\sqrt{1-\mu^2}\left[k-2q\mu\right]q\ F(q,p,\eta)\beta_a(\vec{p})\beta_a(\vec{q})(2\pi)^3\delta(\vec{k}-\vec{p}-\vec{q}),
\end{equation}
where $\mu=\hat{k}\cdot\hat{q}$ and $p=\sqrt{k^2-2kq\mu+q^2}$. For
the expression of anisotropic stress of ordinary matter and radiation,
we refer to, e.g., ref. \cite{Ma:1995ey}. 
Let us define the transfer function for the anisotropic stress of the scalar fields as
\begin{equation}
\label{eq:nlsm_ST}%\eqref{eq:nlsm_ST}
a^2\pi^{\phi}(k,q,\mu,\eta)\equiv \frac{v^2}{2}\sqrt{1-\mu^2}\left[k-2q\mu\right]q\ F(q,p,\eta),
\end{equation}
and the transfer function for $\sigma$ as
\begin{equation}
\label{eq:nlsm_VT}%\eqref{eq:nlsm_VT}
\sigma(k,q,\mu,\eta)=\frac{4\pi}{a^2(\eta)}\frac{v^2}{m_{\rm pl}^2}\int^\eta d\eta'\ a^2(\eta')\sqrt{1-\mu^2}\left[k-2q\mu\right]q\ F(q,p,\eta')/k.
\end{equation}

The fluid equation 
for baryon in the vector mode is given by
\begin{equation}
\label{eq:BV}%\eqref{eq:BV}
\dot{v}_b+{\cal H}v_b=\frac{4\rho_\gamma}{3\rho_b}an_e\sigma_T(v_{\gamma}-v_b),
\end{equation}
where $\rho_\gamma$ and $\rho_b$ are the energy densities of photon and
baryon fluids, respectively, $R=4\rho_{\gamma}/3\rho_b$,  $\sigma_T$ is
the Thomson scattering cross section,  $n_e$ is the electron
number density, and $an_e\sigma_T=\dot{\tau}$ is the opacity of
the Thomson scattering. On the other hand, 
the vector-mode Boltzmann equation of photon fluid expanded in terms of
multipole momenta is given by
\begin{eqnarray}
\label{eq:PV}%\eqref{eq:PV}
\dot{v}_{\gamma}+\frac{1}{8}k\Pi_{\gamma}&=&-\dot{\tau}(v_{\gamma}-v_b),\\
%\end{equation}
%\begin{equation}
\label{eq:Pig}%\eqref{eq:Pig}
\dot{\Pi}_{\gamma}+\frac{8}{5}kI_3-\frac{8}{5}kv_{\gamma}&=&-\dot{\tau}\left(\frac{9}{10}\Pi_{\gamma}-\frac{9}{5}E_2\right)+\frac{8}{5}k\sigma,\\
%\end{equation}
%\begin{equation}
\label{eq:IH}%\eqref{eq:IH}
\dot{I}_l+k\frac{l}{2l+1}\left(\frac{l+2}{l+1}I_{l+1}-I_{l-1}\right)&=&-\dot{\tau}I_l \quad\quad({\rm for~} l\geq3),
\end{eqnarray}
for intensity and
%and the polarization Boltzmann equation as,
\begin{eqnarray}
\label{eq:EH}%\eqref{eq:EH}
\dot{E}_l+\frac{(l+3)(l+2)l(l-1)}{(l+1)^3(2l+1)}kE_{l+1}-\frac{l}{2l+1}kE_{l-1}&=&-\dot{\tau}\left(E_l-\frac{2}{15}\zeta\delta_{l2}\right)+\frac{2}{l(l+1)}kB_l,\qquad\\
\label{eq:BH}%\eqref{eq:BH}
\dot{B}_l+\frac{(l+3)(l+2)l(l-1)}{(l+1)^3(2l+1)}kB_{l+1}-\frac{l}{2l+1}kB_{l-1}&=&-\frac{2}{l(l+1)}kE_l, 
\end{eqnarray}
for polarization.
Here $v_\gamma$ and $\Pi_\gamma$ are the velocity and anisotropic
stress of photons, respectively, $I_l$ is the $l$-th order moment of photons'
distribution, and $E_l$ and $B_l$ are the photons'
polarization moments and $\zeta\equiv 3I_2/4+9E_2/2$ \cite{2004PhRvD..70d3518L}.%\KIQ{Add references}.  
%\KIQ{(KI: give the B-mode polarization equations.)}

As we shall show in section \ref{subsec:MFG},
the relative velocity between the photon and baryon fluids plays the key role in
generation of magnetic fields. 
Since the strength of the coupling between photon and baryon velocities significantly changes
before and after recombination, evolution of the relative velocity and hence the magnetic fields
qualitatively differs between these two epochs. 
Before recombination, the tight-coupling approximation allows us to solve the system of 
equations partially, which we shall see shortly, 
and helps us to interpret numerical results.
On the other hand, after recombination, the system of 
equations is solved completely numerically.

\subsubsection{Tight-coupling approximation}%TCA
\label{subsec:nlsm_TCA}%subsec:nlsm_TCA
%\TSQ{I degraded this part from subsection to subsubsection.}
In the early universe, photon and baryon fluids are tightly coupled
because the opacity of the Thomson scattering $\dot{\tau}$ is very 
large. Therefore we can expand the perturbation equations 
in section \ref{subsec:nlsm_vec} in terms of the 
tight-coupling parameter $k/\dot{\tau}\ll1$. This is called the tight
coupling approximation (TCA). In ordinary analyses without external sources  such as the NLSM, the first
order solution for anisotropic stress of photons $\Pi_g^{(1)}$ and the
second order solution for the relative velocity between the 
photon and 
baryon fluids
$\delta
v^{(2)}=v_\gamma^{(2)}-v_b^{(2)}$ were used \cite{PhysRevD.85.043009}.
However, when there exist NLSM %'s 
scalar fields as an external
source in the linearized Boltzmann system, we find that one
should consider the TCA up to the third order terms proportional to
$\sigma$, as discussed below.

\begin{figure}[!h]
\centering
\includegraphics[width=.75\textwidth]{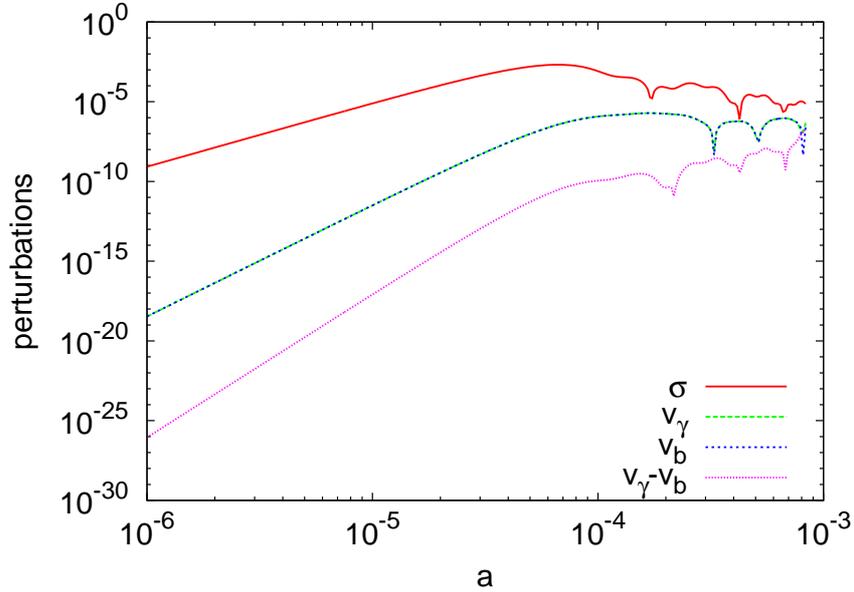}
\caption{\label{fig:nlsm_V} Time-evolutions of the transfer functions. Here 
 $\sigma(k,q,\mu,\eta)$ (red line), $v_\gamma(k,q,\mu,\eta)$ (green), $v_b(k,q,\mu,\eta)$ (blue) and $\delta
 v(k,q,\mu,\eta)$ (magenta) are plotted as functions of the scale factor.
 %Here horizontal axis is the scale factor \TS{and} vertical axis
 %is velocity and vorticity in natural system of  units. 
 We assume $k=q=10^{-1} {\rm Mpc^{-1}}$ and, $\mu=0$. We can see that the condition
 $\sigma\gg v_\gamma$ is almost always satisfied.
 } 
\end{figure}

%When we use
%transfer function $\sigma(k,q,\mu,\eta)$, we can tell $\sigma\gg
%v_\gamma$ , so if we calculate transfer of $\delta v$ in $v_\gamma$'s
%second - order, we need third - order TCA of $\sigma$. 
In the tight coupling expansion, the baryon velocity relative to the
photon velocity is expanded using the tight coupling parameter, i.e.,
$v_\gamma - v_b= 0 + \delta v^{(1)}+\delta v^{(2)}+\dots$, where
$\delta v^{(1)}$ and $\delta v^{(2)}$ are proportional to
$(k/\dot{\tau})$ and $(k/\dot{\tau})^2$, respectively. 
The tight coupling solutions of $\Pi_\gamma$ and $\delta v$ up to the second
order are given by 
\begin{eqnarray}
\label{eq:nlsm_pig3}%\eqref{eq:nlsm_pig3}
\Pi_{\gamma}^{(1)}&=&\frac{32}{15}\left(\frac{k}{\dot{\tau}}\right)(v_{\gamma}^{(0)}+\sigma^{(0)}),
\nonumber \\
\Pi_{\gamma}^{(2)}&=&\frac{32}{15}\left(\frac{k}{\dot{\tau}}\right)(v_{\gamma}^{(1)}+\sigma^{(1)})+\frac{176}{45}\left(\frac{k}{\dot{\tau}}\right)^2\frac{1}{k}\left[\frac{\ddot{\tau}}{\dot{\tau}}(v_{\gamma}^{(0)}+\sigma^{(0)})-(\dot{v}_{\gamma}^{(0)}+\dot{\sigma}^{(0)})\right]
\end{eqnarray}
\begin{eqnarray}
\delta v^{(1)}&=&\left(\frac{k}{\dot{\tau}}\right)\frac{\cal H}{(1+R)k}v_{\gamma}^{(0)}\\
\label{eq:nlsm_dv3}%\eqref{eq:nlsm_dv3}
\delta v^{(2)}&=&\left(\frac{k}{\dot{\tau}}\right)\frac{\cal H}{(1+R)k}v_{\gamma}^{(1)}-\frac{4}{15}\left(\frac{k}{\dot{\tau}}\right)^2\frac{1}{1+R}(v_{\gamma}^{(0)}+\sigma^{(0)})\nonumber\\
& &-\left(\frac{k}{\dot{\tau}}\right)^2\frac{{\cal H}v_{\gamma}^{(0)}}{(1+R)^2k^2}\left(\frac{{\cal H}R}{1+R}+\frac{\dot{\cal H}}{\cal H}+{\cal H}+\frac{\dot{v}_{\gamma}^{(0)}}{v_{\gamma}^{(0)}}-\frac{\ddot{\tau}}{\dot{\tau}}\right)
\end{eqnarray}
It is imporant to
note that $\delta v^{(2)}$ is not necessarily smaller than
$\delta v^{(1)}$ in the NLSM. This is because, in the NLSM, the metric perturbation $\sigma$ is
always much larger than the fluid perturbation variables such as $v_\gamma$ (see figure \ref{fig:nlsm_V} and discussion in section
\ref{sec:D}), and it sometimes
happens that the first order solution proportional to $v_\gamma$ is
smaller than the second order solution proportional to $\sigma$
 %at
%the same order of the third order solution proportional to $\sigma$,
i.e. {$(k/\dot{\tau})v_\gamma\lesssim (k/\dot{\tau})^2 \sigma$}. 
We can  see directly this relation from \eqref{eq:PV}, which implies
\begin{equation}
v_\gamma\sim k\int d\eta\frac{k}{\dot{\tau}}\sigma \sim k\eta \left(\frac{k}{\dot{\tau}}\right)\sigma.
\end{equation}
Therefore, the condition that
{$(k/\dot{\tau})v_\gamma\lesssim (k/\dot{\tau})^2 \sigma$} is satisfied
at least on super-horizon scales, and the slip term is
dominated by the second order terms in the tight coupling
approximation.

In fact, in the numerical calculations, we must eventually switch to
evaluate the
slip term directly from $v_\gamma$ and $v_b$ because the tight coupling
approximation breaks down in late times. Because the slip term is 
dominated by the second order term $\delta
v^{(2)}$ in early times, we must keep solving $v_\gamma$
and $v_b$ accurate enough up to the second order in the tight coupling
approximation, i.e., $v_\gamma^{(2)}$ and $v_b^{(2)}$. In terms of the
tight coupling approximation, the evolution equation of $v_\gamma^{(2)}$,
for instance, is given by
\begin{equation}
\dot{v}_\gamma^{(2)} = -\frac18 k\Pi_\gamma^{(2)}-\dot{\tau}\delta v^{(3)}~.
\end{equation}
This is why we need to consider TCA up to the third order. The slip term at
the third order is given by
\begin{eqnarray}
\delta v^{(3)}& = &\left(\frac{k}{\dot{\tau}}\right)\frac{\cal H}{(1+R)k}v_{\gamma}^{(2)}-\frac{4}{15}\left(\frac{k}{\dot{\tau}}\right)^2\frac{1}{1+R}(v_{\gamma}^{(1)}+\sigma^{(1)})\nonumber\\
& &-\left(\frac{k}{\dot{\tau}}\right)^2\frac{{\cal H}v_{\gamma}^{(1)}}{(1+R)^2k^2}\left(\frac{{\cal H}R}{1+R}+\frac{\dot{\cal H}}{\cal H}+{\cal H}+\frac{\dot{v}_{\gamma}^{(1)}}{v_{\gamma}^{(1)}}-\frac{\ddot{\tau}}{\dot{\tau}}\right)\\
& &+\frac{4}{15}\left(\frac{k}{\dot{\tau}}\right)^3\frac{\cal H}{(1+R)^2k}\sigma^{(0)}\nonumber\\
&&-\frac{2}{45k}\left(\frac{k}{\dot{\tau}}\right)^3\frac{1}{(1+R)^2}\left[(23+11R)\frac{\ddot{\tau}}{\dot{\tau}}\sigma^{(0)}-(17+11R)\dot{\sigma}^{(0)}-\frac{6\sigma^{(0)}{\cal H}R}{1+R}\right]. 
\end{eqnarray}
Here we show only the terms
%we only show the third order terms that are 
proportional to $\sigma$. The condition $\delta v^{(3)}\ll \delta v^{(2)}$ is always valid in
the tight coupling regime.

% Because of this relation, \KI{the slip term $\delta v$ is dominated by 
% the second order terms of TCA in the tight coupling regime.}

% we need to consider the second order terms of TCA when we want to see 
% the contribution of the slip term $\delta v$. To calculate the second order term of the slip terms from Boltzmann equations, we have to see the differential equation of slip term $\delta v$.
%  Arranging \eqref{eq:BV} and \eqref{eq:PV}, we obtain
%  }
% \begin{equation}
% \label{eq:diff_dv}
% \frac{d}{d\eta}\delta v+\frac{1}{8}k\Pi_\gamma-{\cal H}v_b=-\dot{\tau}(1+R)\delta v.
% \end{equation}
% \TS{
% To solve this equation in the accuracy of the second order of TCA, we need the third order expansion of TCA for the right-hand side of this equation. So \eqref{eq:diff_dv} can be rewritten as
% }
% \begin{equation}
% \label{eq:diff_dv}
% \frac{d}{d\eta}\delta v^{(2)}+\frac{1}{8}k\Pi_\gamma^{(2)}-{\cal H}v_b^{(2)}=-\dot{\tau}(1+R)\delta v^{(3)}.
% \end{equation}

%%%%%%something important
%figure

\subsection{Magnetic field generation}%Magnetic field generation
\label{subsec:MFG}%subsec:MFG
%In this sub-section, 
We consider generation of magnetic fields %generation
originated from the relative velocity between the photon and baryon
fluids, $\delta v = v_\gamma-v_b$. Well before recombination, due to the
frequent Thomson 
scattering of photons off electrons, electrons are separated with
protons, and move together with photons. For protons to catch up with
electrons, electric fields are induced and rotation of the induced 
electric fields generates magnetic fields 
via Maxwell equations. 

\begin{figure}[!h]
\centering
\includegraphics[width=.75\textwidth]{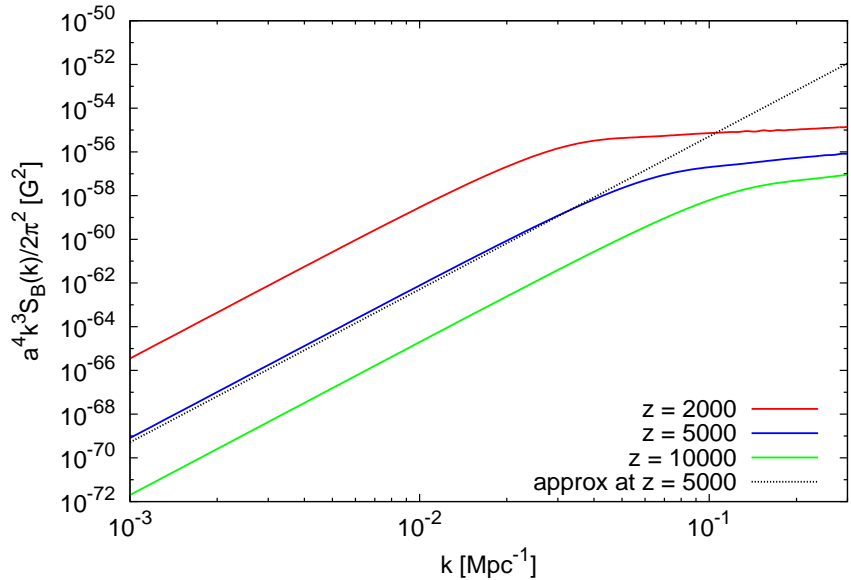}
\caption{\label{fig:nlsm_MFA}  Power spectra of magnetic fields at 
$z=2000$ (red solid line), $z=5000$ (blue solid line) and $z=10000$ (green solid line)  from scalar fields in the NLSM. Here we take the NLSM
 parameters as $v^4/N%$\frac{v^4}{N}
 =10^{-12}m^4_{\rm pl}$. %where $v$ is the VEV
 %and $N$ is the number of the fields. 
%Where horizontal axis
% is wave number $k\ [{\rm Mpc^{-1}}]$, vertical axis is magnetic field's
% power spectrum $a^4\frac{k^3}{2\pi^2}S_B(k)\ [G^2]$. 
The black dotted line is the approximate amplitude which is given by eq. \eqref{eq:nlsm_BPS_approx1}.} 
\end{figure}

The equation for the generation of magnetic fields is given by
\cite{2007astro.ph..1329I}
\begin{equation}
\label{eq:MF_EoM}%\eqref{eq:MF_EoM}
\frac{1}{a}\frac{\rm d}{{\rm d}\eta}(a^2B^i)=\frac{4\sigma_T\rho_{\gamma}a}{3e}\epsilon^{ijk}\partial_k(v_{\gamma j}-v_{bj}),
\end{equation}
where $e$ is the elementary charge
and $\epsilon^{ijk}$ is the Levi-Civita tensor. The appearance of the rotation of $\delta v$ in eq. \eqref{eq:MF_EoM} clearly shows that 
only the vector-mode part of $\delta v$ can contribute to magnetic fields.
By integrating eq. \eqref{eq:MF_EoM} in Fourier space, we obtain
\begin{eqnarray}
\label{eq:MF_IS}%\eqref{eq:MF_IS}
a^4B^i(\vec{k},\eta)B_i^*(\vec{k}',\eta)&=&\left(\frac{4\sigma_T}{3e}\right)^2(\delta^{jl}\delta^{km}-\delta^{jm}\delta^{kl})k_kk'_m\int _0^{\eta} {\rm d}\eta'a^2(\eta')\rho_{\gamma}(\eta')\delta v_j(\vec{k},\eta')\nonumber\\
& &\times\int _0^{\eta} {\rm d}\eta''a^2(\eta'')\rho_{\gamma}(\eta'')\delta v_l^*(\vec{k}',\eta'').
\end{eqnarray}
Next we take an ensemble average of this expression
over the initial configuration of the NLSM scalar fields $\beta_a(\vec k)$.
The ensemble average of the relative velocity can be calculated using
the transfer function $\delta v (k,q,\mu,\eta)$ and the NLSM's initial power
spectrum ${\cal P}_{\rm ini}^N$ defined in %the 
appendix \ref{subsec:AA} as
\begin{eqnarray}
\label{eq:MF_nlsm_IS}%\eqref{eq:MF_nlsm_IS}
\left<
\delta v_j(\vec{k},\eta')\delta v^*_l(\vec{k}',\eta'')\right>
&=&P_{jl}(\hat{k})\frac{{\cal P}_{\rm ini}^N}{{2\pi^2}}(2\pi)^3\delta(\vec{k}-\vec{k}')\nonumber\\
& &\times\int dq\ q^2\int d\mu\ \delta v (k,q,\mu,\eta') \delta v (k,q,\mu,\eta''),\\\nonumber\\
\label{eq:Projection}%\eqref{eq:Projection}
P_{jl}(\hat{k})&=&\delta_{jl}-\hat{k}_j\hat{k}_l.
\end{eqnarray}
The correlation function of magnetic fields is then obtained as
%Then the ensemble average of magnetic fields is expressed as,
\begin{equation}
%\left<
\left<
B^i(\vec{k},\eta)B_i^*(\vec{k}',\eta)
\right>
%\right>
=(2\pi)^3S_B(k,\eta)\delta(\vec{k}-\vec{k}'),
\end{equation}
where
\begin{equation}
\label{eq:nlsm_BPS}%\eqref{eq:nlsm_BPS}
a^4(\eta)\frac{k^3}{2\pi^2}S_B(k,\eta)=\frac{k^3}{2\pi^2}\left(\frac{4\sigma_T}{3e}\right)^2\frac{{\cal P}_{\rm ini}^N}{{\pi^2}}k^2\int dq\ q^2\int_{-1}^1d\mu\ \left[\int _0^{\eta} {\rm d}\eta'\ a^2(\eta')\rho_{\gamma}(\eta')\delta v(k,q,\mu,\eta')\right]^2.
\end{equation}
We calculate eq.~\eqref{eq:nlsm_BPS} numerically
and the power spectra of magnetic fields at several redshifts are depicted
%show the power spectra at some redshifts 
in figure \ref{fig:nlsm_MFA}. %Where we assume scalar field  following NLSM as
%only originated seed of magnetic field, vorticity from scalar field has
%the strongest contribution to magnetic field. \KIQ{(KI: I could not get
%the meaning of this sentence.)}

\section{Analytical Understanding}%Disscassion
\label{sec:D}%sec:D
Let us try to understand the results obtained in
the previous section analytically.
% we discuss about the power spectrum of magnetic field. 
%We could get the power spectrum of magnetic field in section
%\ref{subsec:MFG}. 
In our numerical calculations, we consider scalar fields
following the NLSM as the only source of vector mode cosmological
perturbations. Therefore we assume that there are no vector mode
perturbations at $\eta \to 0$. In this setup, 
vorticity $\sigma$ evolves first with the scalar fields as an
external source (eq. \eqref{eq:nlsm_AS}), and it induces photons'
anisotropic stress $\Pi_\gamma$. Then $\Pi_\gamma$ leads to the photon
velocity $v_\gamma$ and it propagates to the baryon velocity
$v_b$. Because the induced anisotropic stress $\Pi_\gamma$ is
suppressed by a factor of the tight coupling parameter $k/\dot{\tau}$
due to the frequent Thomson scattering (see eq.~\eqref{eq:nlsm_pig3}), the
velocities induced from the anisotropic stress are much smaller than
$\sigma$, in other words, the condition that 
$\sigma\gg v_\gamma\sim v_b$  is valid
at least in the tight coupling era. Therefore, we can find that the dominant
term in eq. \eqref{eq:nlsm_dv3} is
\begin{equation}
\label{eq:nlsm_dvA}%\eqref{eq:nlsm_dvA}
\delta v(k,q,\mu,\eta)\approx -\frac{4}{15}\left(\frac{k}{\dot{\tau}}\right)^2\frac{1}{1+R}\sigma(k,q,\mu,\eta). 
\end{equation}
% vorticity $\sigma$ evolves following \eqref{eq:EoMS} and its evolution
% propagates to velocities via Boltzmann equation \eqref{eq:PV} and
% \eqref{eq:Pig} with tight coupling parameter $k/\dot{\tau}$ (see
% \eqref{eq:nlsm_pig3}), So we can calculate magnetic field with
% assumption of \eqref{eq:nlsm_dvA}.And magnetic field's power spectrum
% propers to vorticity $\sigma$'s power spectrum, so it is proportional to
% $v^4/N$ too.  Here we calculate $k$ dependence of magnetic field' power
% spectrum originated in NLSM. To estimate power spectrum of magnetic
% field, because $\sigma \gg v_\gamma$ is to be practical in transfer
% function, we can assume as
%We can calculate $k$ dependence of \eqref{eq:nlsm_BPS} using
%\eqref{eq:nlsm_dvA}, but we need notation of transfer function
%$\sigma(k,q,\mu,\eta)$. 
To calculate $\sigma(k,q,\mu,\eta)$, we introduce approximations
for the Bessel function in eq.~\eqref{eq:nlsm_sol} which are given by 
\begin{eqnarray}
\label{eq:bessel}%\eqref{eq:bessel}
\frac{J_\nu(x)}{x^\nu}\approx
\begin{cases}
\frac{1}{2^\nu\Gamma(\nu+1)} & ({\rm for~} x\ll 1) \\
\frac{1}{x^\nu}\sqrt{\frac{2}{\pi x}}{\rm cos}\left(x-\frac{2\nu+1}{4}\pi\right) & ({\rm for~} x\gg1)
\end{cases}.
\end{eqnarray}
Using these approximations, we can calculate super-horizon
$(k\eta\ll1)$ and sub-horizon $(k\eta\gg1)$ solutions in the radiation
and matter dominated eras as we shall show below.

\subsection{Super-horizon}%Super-horizon
\label{subsec:D_super-horizon}%\ref{subsec:D_super-horizon}
On 
%At 
super-horizon scales, the wavenumber of fluctuations $k$ 
is smaller than the inverse of the horizon scale, i.e.,
%can be denoted as 
$k\eta\ll 1$. However the wavenumber $q$ in
eq. \eqref{eq:nlsm_BPS} which comes from the convolution integral
does not necessarily satisfy $q\eta\ll 1$.
%we need to know how we can denote 
%wave number of scalar field $p,\ q$.  
We
%Fortunately, we 
know from eq. \eqref{eq:nlsm_ST} and eq. \eqref{eq:bessel}
that $\sigma(k,q,\mu,\eta)$ as a function of $q$ decays as
$\sigma \propto q^{-2\nu+1}$ for $q\gg \eta^{-1}$, and we could expect that the
bulk of the $q$-integration comes from the range of $q\lesssim \eta^{-1}$.
%about $k\ll q\sim p\sim 1/\eta$. So we should assume the range of
%integration of scalar field's wave number $q$ as $k\ll p\sim q <
%1/\eta$. 
%Here we can denote \eqref{eq:nlsm_BPS}, 
In this case we can express eq.~\eqref{eq:nlsm_BPS} as
\begin{equation}
\label{eq:nlsm_BPS_super-horizon}%\eqref{eq:nlsm_BPS_super-horizon}
a^4\frac{k^3}{2\pi^2}S_B(k)\propto k^5\eta^{-3}\left[\eta a^{-2}\delta v(k,q=1/\eta,\mu,\eta)\right]^2.
\end{equation}
For $k \ll q\sim p\sim 1/\eta$, we can find the $k$-dependence of
$\sigma$ from eq.~\eqref{eq:nlsm_VT} as, 
\begin{equation}
\label{eq:nlsm_sigma_k_dependence_super-horizon}%\eqref{eq:nlsm_sigma_k_dependence_super-horizon}
\sigma\propto \frac{\eta^2}{k}.
\end{equation}
Then, substituting eq. \eqref{eq:nlsm_dvA} and
eq. \eqref{eq:nlsm_sigma_k_dependence_super-horizon} into
eq. \eqref{eq:nlsm_BPS_super-horizon}, we obtain
\begin{equation}
\label{eq:nlsm_bps_k_super-horizon}%\eqref{eq:nlsm_bps_k_super-horizon}
a^4\frac{k^3}{2\pi^2}S_B(k)\propto k^5\left[\frac{k^2}{k}\right]^2\propto k^7.
\end{equation}
We can see this power law tail on large scales in figure \ref{fig:nlsm_MFA}.
The spectral index is same as that of the magnetic fields generated from
second order density perturbations \cite{2007astro.ph..1329I}, but slightly different from the one
obtained in the Einstein-aether gravity model, where
$\sqrt{\left<B^2_{EA}(k)\right>}\propto k^4$ \cite{Saga:2013glg}.

%In section \ref{subsubsec:super-horizon}, we say we can estimate $p\sim q\sim 1/\eta$, about horizon scale. So we can  integrate  
% \eqref{eq:nlsm_BPS} as \eqref{eq:nlsm_bps_k_super-horizon}, we can say $a^4\frac{k^3}{2\pi^2}S_B\propto k^7$ as analytic solution. In simulation, we can see the same power law at super-horizon scale  (see Figure.\ref{fig:nlsm_MFA}). So we can say vorticity $\sigma$ is the most dominant term in  contributions to magnetic field and horizon scale of scalar fields affect  the most. 
 
\subsection{Sub-horizon}%Sub-horizon
On sub-horizon scales, i.e. $k\eta\gg1$, the situation changes in a more complicated way.
Because of the condition that $p\eta=\sqrt{k^2-2kq\mu+q^2}\eta\gg1$, we
must take special care of rapid oscillations of the Bessel functions 
in eqs.\eqref{eq:nlsm_VT} and \eqref{eq:nlsm_BPS}. In order to
manipulate the equations analytically,  
we divide the interval of integration with wavenumber $q$ into three
regions: 
(i) $\ k\sim q\gg\eta^{-1}$; $q$ is on sub-horizon scale,  (ii) $\ q\ll
\eta^{-1}$; $q$ is on super-horizon scale, and (iii) 
$\ \eta^{-1}<q<\alpha \eta^{-1}$; $q$ is nearly on the horizon scale (with $\alpha$ being
${\cal O}(1)$ constant).
Considering contributions from each interval, we estimate the power
spectrum of magnetic fields.
\subsubsection{case (i) $\ k\sim q\gg\eta^{-1}$}
In this case, because the conditions that 
$p\eta\gg 1$ and $\ q\eta\gg1$ are satisfied from the conservation of the momentum,
the source function of the vorticity $\sigma$ has decayed
away. Without sources, the vorticity also decays and therefore the contribution from this part is negligible.
%Once we write the power spectrum from this region, $B^2(k)\propto k^{0}$.\\
\subsubsection{case (ii)$\ q\ll \eta^{-1}$}
In this case, $p\sim k$ and the source of the vorticity is growing. Using the
approximation of eq.~\eqref{eq:bessel}, and assuming the radiation
dominated era ($\nu=2$), we can evaluate eq.~\eqref{eq:nlsm_VT} as
\begin{eqnarray}
\label{eq:sigma_super_horizon_int}
\sigma&\propto&\frac{1}{a^2}\int_0^{\eta'} d\eta''\ \eta''^5q(1-2q\mu/k)\frac{J_\nu(k\eta'')}{(k\eta'')^\nu},\nonumber\\
&\propto&\eta'^{-2}k^{-6}q(k\eta')^3J_3(k\eta').
\end{eqnarray}
Here we use the fact that $1-2q\mu/k\simeq1$ and the formula $\int dx\
x^{n+1}J_{n}(x)=x^{n+1}J_{n+1}(x)$, and ignore the factor
$\sqrt{1-\mu^2}$. Using the above formula again and eq.~\eqref{eq:nlsm_dvA}, we obtain
\begin{equation}
\label{eq:dv_super_horizon_int}
\int_0^{\eta} d\eta'\ \rho_\gamma a^2\delta v\propto k^{-6}q(k\eta)^4J_4(k\eta).
\end{equation}
Substituting eq.~\eqref{eq:dv_super_horizon_int} into eq.~\eqref{eq:nlsm_BPS} and using eq.~\eqref{eq:bessel},
the spectrum can be written as
\begin{eqnarray}
a^4\frac{k^3}{2\pi^2}S_B(k)&\propto& k^5\int_0^{1/\eta} dq\ q^4\left[k^{-6}(k\eta)^4J_4(k\eta)\right]^2\nonumber\\
&\propto& k^0~.
\end{eqnarray}

\subsubsection{case (iii) $\ \eta^{-1}<q<\alpha \eta^{-1}$}
In this range, we need to treat the source carefully. First, assuming the
radiation dominated era, let us
divide the $\eta^\prime$ integral in eq.~\eqref{eq:nlsm_BPS} as 
\begin{equation}
\int_0^\eta a^2 (\eta^\prime)\rho_\gamma(\eta^\prime) \delta v\ d\eta' \propto
 \int^\eta_0\eta' \sigma(\eta^\prime) d\eta^\prime =\int_0^{q^{-1}}
 \eta'\sigma_{\eta^\prime < q^{-1}}(\eta^\prime) d\eta^\prime + \int_{q^{-1}}^\eta
\eta' \sigma_{\eta^\prime > q^{-1}}(\eta^\prime) d\eta^\prime~.
\end{equation}
The integrand of the first term in the above equation,
$\sigma_{\eta^\prime < q^{-1}}(\eta^\prime)$, is given by eq.~\eqref{eq:sigma_super_horizon_int}, and
the integration leads to the term proportional to $k^{-6}q(k/q)^4
J_4(k/q)$. That of the second term is calculated as
\begin{eqnarray}
 \sigma_{\eta^\prime > q^{-1}}(\eta^\prime) \simeq &&
  \frac{1}{a^2}\int_0^{q^{-1}}d\eta^{\prime\prime}\eta''^5q(1-2q\mu/k)\frac{J_\nu(k\eta'')}{(k\eta'')^\nu}
  \nonumber \\
+&&
  \frac{1}{a^2}\int_{q^{-1}}^{\eta^\prime} d\eta^{\prime\prime}\eta''^5q(1-2q\mu/k)\frac{J_\nu(k\eta'')}{(k\eta'')^\nu} \frac{J_\nu(q\eta'')}{(q\eta'')^\nu} 
\nonumber \\
\propto &&\eta'^{-2}k^{-6}[q(k/q)^3J_3(k/q)+k^{2}q^{-1}(k\eta')(J_3(k\eta')J_2(q\eta')+{\cal O}(q/k))]
\end{eqnarray}
where we used $J_\nu(q\eta'')/(q\eta'')^\nu\simeq {\cal O}(1)$ for
$q\eta''\ll 1$, and omitted the constant factor of ${\cal O}(1)$.

Then the integration of $\delta v$ can be calculated, by integrating
by parts, as
\begin{eqnarray}
\label{eq:dv_sub_horizon_int}
\int_0^{\eta} d\eta'\ \rho_\gamma a^2\delta v&\propto& \left[k^{-6}q(k/q)^4J_4(k/q)
+k^{-4}q(k/q)^3J_3(k/q)(\eta^2-q^{-2})\right.\nonumber\\
&+&\left.k^{-4}q^{-1}(k\eta)^2J_4(k\eta)J_2(q\eta)\right].
\end{eqnarray}
Substituting the above equation into eq.\eqref{eq:nlsm_BPS} and ignoring
the cross terms, we obtain,
\begin{eqnarray}
a^4\frac{k^3}{2\pi^2}S_B(k)&\propto& k^5\int_{1/\eta}^{\alpha/\eta} dq\ q^2\left[k^{-12}q^2(k/q)^8J_4^2(k/q)\right.\nonumber\\
&+&\left.k^{-8}q^2(k/q)^6J_3^2(k/q)(\eta^2-q^{-2})^2+k^{-8}q^{-2}(k\eta)^4J_4^2(k\eta)J_2^2(q\eta)\right].
\end{eqnarray}
Integrating with $q$, we find that the first, second, and the third
terms in the above equation give terms  $\propto k^0$, $\propto k^1$, and $\propto k^0$, respectively.
Taking
these terms together, we can find the $k$ dependence of the magnetic
field spectrum as 
\begin{equation}
a^4\frac{k^3}{2\pi^2}S_B(k)\propto k[1+{\cal O}(1/k\eta)].
\end{equation}
Therefore, in the radiation dominated era ($\nu=2$), we
find $a^4k^3S_B(k)/2\pi^2\propto k$, which is confirmed
in our numerical calculation.
%Above discussions are 
%only order estimation, and to calculate correct values we need to 
%do integrations  making use of stepwise development.
%So it's too hard to do by hand calculation, 
%here we read 
%the amplitude and extrapolate the spectrum, 
Finally, by reading off the numerical amplitude from the result of our
numerical calculation we find the power spectrum of magnetic field on
sub-horizon scales as 
\begin{equation}
a^4B^2\sim
 \frac{10^{-44}}{(1+z)^3}\frac{1}{N}\left(\frac{v}{10^{-3}m_{pl}}\right)^4\left(\frac{k}{{\rm
					    Mpc}^{-1}}\right)^{1}~{[{\rm
 G}^2]}
\end{equation}
Similarly, we find $a^4k^3S_B(k)/2\pi^2\propto k^{-1}$ in the matter
dominated era ($\nu=3$).

In figure \ref{fig:nlsm_MFA}, we find that the spectrum shows
$k^3 S_B \propto k$ on small scales (say, $k\gtrsim 0.1$ Mpc$^{-1}$) in
the radiation dominated era. At $z=2000$ (the red solid line), the
spectrum on sub-horizon scales shows the $k$ dependence between the fully
radiation dominated ($\propto k$) and matter dominated ones ($\propto
k^{-1}$). 
On much smaller scales ($k\gg
1$ Mpc$^{-1}$) and in the matter dominated era, we expect that the 
spectrum of magnetic fields should be proportional to $k$ because on those scales the
source of vector perturbations has already decayed away %is absent 
and the magnetic fields just
decay adiabatically after their creation deep in the radiation dominated
era. From 
the fact that the generation mechanism is based
on the mass difference between positively charged particles (protons)
and negatively charged particles (electrons) and the small velocity slip
between these particles, we expect that the
spectrum continues up to the horizon scale at the epoch of $e^\pm$
annihilation $k \sim 10^{10}$ Mpc$^{-1}$ and a cutoff at that scale \cite{PhysRevD.85.043009}.

\subsection{Approximation at super-horizon scale}
\label{subsubsec:approximation}%\ref{subsubsec:approximation}
On super-horizon scales and in the
radiation dominated era,  we can estimate 
not only the shape of the spectrum but also the
amplitude of magnetic fields approximately. 
On super-horizon scales
%In this case we can denote
the power spectrum of magnetic fields is given by %as 
\begin{equation}
\label{eq:nlsm_BPS_approx0}%\eqref{eq:nlsm_BPS_approx}
a^4(\eta)\frac{k^3}{2\pi^2}S_B(k,\eta)=\frac{k^3}{2\pi^2}\left(\frac{4\sigma_T}{3e}\right)^2\frac{{\cal P}_{\rm ini}^N}{{\pi^2}}k^2\int_0^{1/\eta} dq\ q^2\int_{-1}^1d\mu\ \left[\int _0^{\eta} {\rm d}\eta'\ a^2(\eta')\rho_{\gamma}(\eta')\delta v(k,q,\mu,\eta')\right]^2.
\end{equation}
%We can denote $\delta v(k,q=1/\eta,\mu,\eta)$ too, substituting
%\eqref{eq:bessel} and \eqref{eq:nlsm_ST} to \eqref{eq:EoMS}, we can
%get,
Substituting eq.
\eqref{eq:bessel} and eq.  \eqref{eq:nlsm_ST} to eq. \eqref{eq:EoMS} we get,
\begin{equation}
\label{eq:nlsm_sigma_approx}%\eqref{eq:nlsm_sigma_approx}
\sigma(k,q,\mu,\eta)\simeq-\frac{\pi A_\nu}{48}\sqrt{1-\mu^2}\mu\left(\frac{v}{m_{pl}}\right)^2\eta^4\frac{q^2}{k},
\end{equation}
where we assume $k\ll q\sim 1/\eta$. Using the above expression, we
can write %calculate 
the velocity slip as
\begin{equation}
\label{eq:nlsm_dv_approx}%\eqref{eq:nlsm_dv_approx}
\delta v(k,q,\mu,\eta)\approx \frac{\pi A_\nu}{180}R^{-1}\dot{\tau}^{-2}\sqrt{1-\mu^2}\mu\left(\frac{v}{m_{pl}}\right)^2\eta^4q^2k. 
\end{equation}
Substituting  eq.\eqref{eq:nlsm_dv_approx} into eq.\eqref{eq:nlsm_BPS_approx0}, we obtain
\begin{equation}
\label{eq:nlsm_BPS_approx1}%\eqref{eq:nlsm_BPS_approx}
a^4(z)\frac{k^3}{2\pi^2}S_B(k,z)\sim\frac{10^{-3}}{(1+z)^9}\frac{1}{N}\left(\frac{v}{m_{pl}}\right)^4\left(\frac{k}{\rm Mpc^{-1}}\right)^7 \ {\rm [G^2]},
\end{equation}
in unit of Gauss at redshift $z\gg z_{\rm eq} \approx 3300$  \cite{Ade:2013zuv}. 
This analytic power is plotted in figure\ref{fig:nlsm_MFA} to make a comparison with 
numerical results.

%Substituting all values at $z=1200$ and assuming flat spectrum, we can get 
%\begin{equation}
%\label{eq:nlsm_BPS_approx_const}%\eqref{eq:nlsm_BPS_approx_const}
%a^4(\eta)\frac{k^3}{2\pi^2}S_B(k,\eta)\sim10^{-41\sim40}\left(\frac{v}{m_{pl}}\right)^4\ \left[{\rm G^2}\right],
%\end{equation}
%in the unit of gauss.

%We can get constraint on vacuum expected value by using \eqref{eq:nlsm_BPS_approx_const} and lower bounds of intergalactic magnetic fields $10^{-20}-10^{-16}{\rm G}$. So we can tell $10^{-20}-10^{-16}{\rm G} < k^3S_B(k)/2\pi^2$, and $(10^{-3}-10^{-1})m_{\rm pl}< v$. 

%So we need to know dealing of $q,\ p$ too. In section \ref{subsubsec:sub-horizon}, we see 
%$q\sim p\sim k$. Using this relation, we can estimate magnetic field's power spectrum as \eqref{eq:nlsm_bps_k_sub-horizon}, so we can get scale invariant spectrum $a^4\frac{k^3}{2\pi^2}S_B\propto k^0$ in matter dominated era. We estimate this scale invariant power spectrum approximately in section \ref{subsubsec:approximation} and we draw the approximation in Figure.\ref{fig:nlsm_MFA} with result of simulation. In Figure.\ref{fig:nlsm_MFA}, we can see scale invariant spectrum in small scale and we can tell that simulation consists with approximation. So this result indicates that the assumption of wave number  $q\sim p\sim k$ is right at least in tight coupling era.

\subsection{After recombination}

In the above sections, we discussed magnetic field generation in the
era when the tight coupling approximation is valid. 
Here we consider the era after the tight coupling approximation breaks down.
In particular, we estimate when the magnetic fields 
become source-free on super-horizon scales, by considering the 
time evolution of the source function 
of magnetic fields $S(\eta)=\rho_\gamma a^2 \delta v q^{3/2}$, which satisfies
$B^2\propto[\int d\eta S]^2$. 

\begin{figure}[!h]
\centering
\includegraphics[width=.75\textwidth]{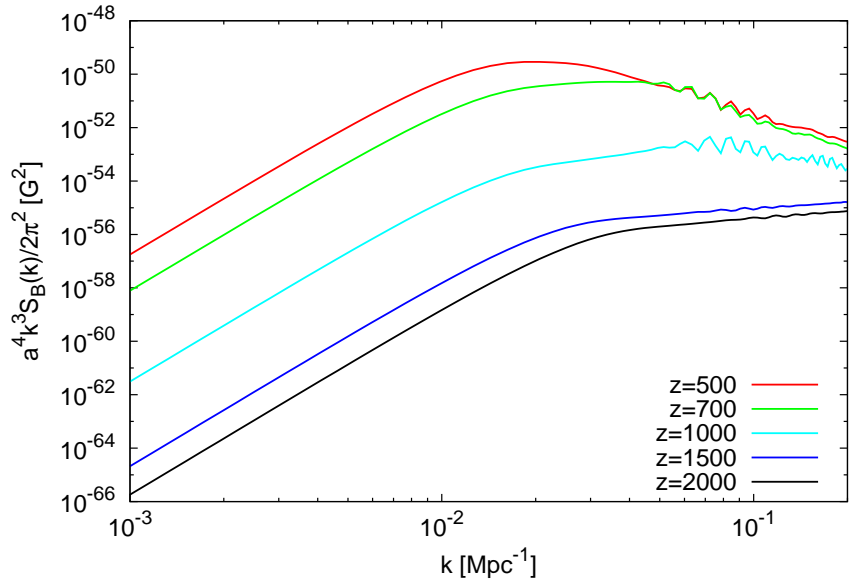}
\caption{\label{fig:nlsm_spec_z10}  Power spectra of magnetic fields at 
$z=2000$ (black solid line), $1500$ (green solid line), $1000$ (cyan solid line), $700$ (green solid line), and $500$ (red solid line)  from scalar fields in the NLSM. Here we take the NLSM
 parameters as $v^4/N%$\frac{v^4}{N}
 =10^{-12}m^4_{\rm pl}$. %where $v$ is the VEV
 %and $N$ is the number of the fields. 
%Where horizontal axis
% is wave number $k\ [{\rm Mpc^{-1}}]$, vertical axis is magnetic field's
% power spectrum $a^4\frac{k^3}{2\pi^2}S_B(k)\ [G^2]$. 
The evolution of the spectrum comes to an end by $z=300$.} 
\end{figure}

On super-horizon scales $(k\ll{\cal H})$, the scalar fields with
wavenumbers $q\sim {\cal H}$ have the biggest contribution to the source
of vector perturbations and hence the magnetic fields. 
Thus we can set $q\sim 1/\eta$ in investigating the behavior of the source. 
When the tight coupling approximation is valid, i.e. $z\gg z_{\rm
rec}$, we can estimate the time evolution of $\delta v$ as 
\begin{equation}
\label{vg_tca}
\delta v\ \propto\ k^2\frac{\eta}{\dot{\tau}}\sigma\ \propto\ \eta^7,
\end{equation}
from eq. \eqref{eq:PV} and eq. \eqref{eq:Pig}. Then 
after recombination, we can estimate $\delta v\approx v_\gamma$ from the 
same equations as
\begin{equation}
\delta v \propto\ k^2\eta^2\sigma\ \propto\ \eta^4.
\end{equation}
Using these relations, the time evolution of the source
term of magnetic fields before and after recombination can be derived as
\begin{equation}
\label{src_t_evol}
S(\eta)=\rho_\gamma a^2v_\gamma q^{3/2}\propto\left\{ \begin{array}{ll}
\eta^{3/2} & ({\rm before\ recombination})\\
\eta^{-3/2} & ({\rm after\ recombination})\end{array}\right..
\end{equation} 
Therefore, the evolution of magnetic fields becomes source-free after
recombination. In fact, during recombination, $\delta v$ is considerably
enhanced and significant amount of magnetic fields is produced by the
end of recombination $z\gtrsim 300$. For $z\lesssim 300$, the magnetic
fields simply decay adiabatically. The spectrum of magnetic fields after
recombination is depicted in figure \ref{fig:nlsm_spec_z10}.

\section{Conclusion}
In this paper, we consider magnetic field generation from
self-ordering scalar fields that follow the NLSM.  We find that to reliably
estimate the magnetic fields, one needs to expand
the Boltzmann equations up to the third order terms in the tight
coupling approximation. This is because the vorticity $\sigma$ is very
large so that {${\cal O}\left((k/\dot{\tau})v_\gamma\right)\sim {\cal
O}\left((k/\dot{\tau})^3 \sigma\right)$}
 in the tight coupling era (see
figure.\ref{fig:nlsm_V}) if the anisotropic stress of the scalar
fields eq. \eqref{eq:nlsm_AS} is an external source. By smoothly connecting the tight coupling
solutions to the numerical ones we obtain the full magnetic field spectrum
in the radiation dominated era and the matter
dominated era, with an analytic interpretation of the results.
In so doing, 
we see that the scalar
fields with wavenumber $q\sim 1/\eta$ are the main source for both 
super-horizon ($k\eta\ll1$) and sub-horizon ($k\eta\gg1$) magnetic fields. 
By extrapolating our numerical result toward smaller scales analytically, we find %can expect 
$B\sim 10^{-22}\left(v/10^{-3}m_{pl}\right)^2\left(1+z\right)^{1/2}\left(k/{\rm Mpc}^{-1}\right)^{1/2}/\sqrt{N}$ Gauss at $k\gtrsim 1$ Mpc$^{-1}$ and
$z\gtrsim z_{\rm rec}$. %\KIQ{Comment on current constraints on $v$.} 
This might serve as a seed of large scale magnetic
fields in the present universe.

% At first, we introduced self-ordering
% scalar fields following NLSM and saw its solution in
% section.\ref{sec:nlsm}. Second, we reviewed scalar vector tensor
% decomposition and tight-coupling approximation, and we could calculate
% magnetic field's power spectrum in section \ref{sec:magnetic}. Finally,
% we discussed about power spectrum of magnetic field in section
% \ref{sec:D}. Here, we summarize important points of physics in this
% magnetic field.

%  In simulation, we can see power law and scale
% invariant power spectrum of magnetic fields, and we can see its power
% \eqref{eq:nlsm_bps_k_super-horizon} and amplitude
% \eqref{eq:nlsm_bps_k_sub-horizon} analytically, so we can tell horizon

\label{sec:C}%sec:C

\section*{Acknowledgement}
This work is supported by Grant-in-Aid for Scientific Research No.25287057 (N.S.) and 24340048 (K.I.).
T.S. is supported by the Academy of Finland grant 1263714.
\appendix
\section{Appendix}%Appendix
\subsection{Appendix A : Power spectrum in the NLSM}%Appendix A
\label{subsec:AA}%subsec:AA
Let us consider the power spectrum of some variable $X(\vec{k})$ and
$Y(\vec{k})$ which
is generated by scalar fields following the NLSM. At first, let us write
$X(\vec{k})$ using its transfer function $F_X(q,p)$ and $F_Y(q,p)$ as 
\begin{equation}
\label{eq:nlsm_X}%\eqref{eq:nlsm_X}
X(\vec{k})=\int \frac{{\rm d}^3p}{(2\pi)^3}\frac{{\rm d}^3q}{(2\pi)^3}F_X(q,p)\beta_s(\vec{p})\beta_s(\vec{q})(2\pi)^3\delta(\vec{k}-\vec{p}-\vec{q}),
\end{equation}
and a similar expression for $Y(\vec{k})$. The (cross) power spectrum of
$X$ and $Y$ is defined by
\begin{equation}
\label{eq:nlsm_P}%\eqref{eq:nlsm_P}
\left<X(\vec{k})Y^*(\vec{k}')\right>\equiv (2\pi)^3{\cal P}_{XY}(k)\ \delta(\vec{k}-\vec{k}').
\end{equation}
To calculate the power spectrum, we need to calculate four point
correlation function of scalar fields. This is given by  
\begin{eqnarray}
\label{eq:nlsm_IPS}%\eqref{eq:nlsm_IPS}
\left<\beta_a(\vec{q}_1)\beta_a(\vec{p}_1)\beta_b^*(\vec{q}_2)\beta^*_b(\vec{p}_2)\right>&=&\left<\beta_a(\vec{q}_1)\beta_a(\vec{p}_1)\right>\left<\beta^*_b(\vec{q}_2)\beta^*_b(\vec{p}_2)\right>\nonumber\\
& &+\left<\beta_a(\vec{q}_1)\beta_b^*(\vec{q}_2)\right>\left<\beta_a(\vec{p}_1)\beta_b^*(\vec{p}_2)\right>\nonumber\\
& &+\left<\beta_a(\vec{q}_1)\beta_b^*(\vec{p}_2)\right>\left<\beta_a(\vec{p}_1)\beta_b^*(\vec{q}_2)\right>\nonumber\\
&=&(6\pi^2)^2\ (2\pi)^3\delta(\vec{p}_1+\vec{q}_1)\ (2\pi)^3\delta(\vec{p}_2+\vec{q}_2)\nonumber\\
& &+\frac{(6\pi^2)^2}{N}\ (2\pi)^3\delta(\vec{q}_1-\vec{q}_2)\ (2\pi)^3\delta(\vec{p}_1-\vec{p}_2)\nonumber\\
& &+\frac{(6\pi^2)^2}{N}\ (2\pi)^3\delta(\vec{q}_1-\vec{p}_2)\ (2\pi)^3\delta(\vec{p}_1-\vec{q}_2).
\end{eqnarray}
Here, $\frac{(6\pi^2)^2}{N}={\cal P}_{\rm ini}^N$ is the initial power
spectrum of the scalar fields. Substituting eq. \eqref{eq:nlsm_IPS} to
eqs. \eqref{eq:nlsm_X} and \eqref{eq:nlsm_P}, and using 
$F(q,p)=F(p,q)$, we calculate the cross correlation as 
\begin{eqnarray}
\label{eq:nlsm_PSC}%\eqref{eq:nlsm_PSC}
\left<X(\vec{k})Y^*(\vec{k}')\right>&=&\int \frac{d^3q_1}{(2\pi)^3}\frac{d^3p_1}{(2\pi)^3}\frac{d^3q_2}{(2\pi)^3}\frac{d^3p_2}{(2\pi)^3}F_{X}(q1,p_1)F_{Y}^*(q_2,p_2)\left<\beta_a(\vec{q}_1)\beta_a(\vec{p}_1)\beta_b^*(\vec{q}_2)\beta_b^*(\vec{p}_2)\right>\nonumber\\
& &\times(2\pi)^3\delta(\vec{k}-\vec{p}_1-\vec{q}_1)\ (2\pi)^3\delta(\vec{k}'-\vec{p}_2-\vec{q}_2)\nonumber\\
&=&\int \frac{d^3q_1}{(2\pi)^3}\frac{d^3p_1}{(2\pi)^3}\frac{d^3q_2}{(2\pi)^3}\frac{d^3p_2}{(2\pi)^3}F_{X}(q_1,p_1)F_{Y}^*(q_2,p_2)\nonumber\\
& &\left[\frac{(6\pi^2)^2}{N}\ (2\pi)^3\delta(\vec{q}_1-\vec{q}_2)\ (2\pi)^3\delta(\vec{p}_1-\vec{p}_2)+\frac{(6\pi^2)^2}{N}\ (2\pi)^3\delta(\vec{q}_1-\vec{p}_2)\ (2\pi)^3\delta(\vec{p}_1-\vec{q}_2)\right]\nonumber\\
& &\times(2\pi)^3\delta(\vec{k}-\vec{p}_1-\vec{q}_1)\ (2\pi)^3\delta(\vec{k}'-\vec{p}_2-\vec{q}_2)\nonumber\\
&=&{\cal P}_{\rm ini}^N(2\pi)^3\delta(\vec{k}-\vec{k}')\nonumber\\
& &\times\int \frac{d^3q}{(2\pi)^3}\left[F_X(q,|\vec{k}-\vec{q}|)F_Y^*(q,|\vec{k}-\vec{q}|)+F_X(|\vec{k}-\vec{q}|,q)F_Y^*(q,|\vec{k}-\vec{q}|)\right]\nonumber\\
&=&2{\cal P}_{\rm ini}^N(2\pi)^3\delta(\vec{k}-\vec{k}')\int \frac{d^3q}{(2\pi)^3}F_X(q,|\vec{k}-\vec{q}|)F_Y^*(q,|\vec{k}-\vec{q}|)~.
\end{eqnarray}
Therefore the power spectrum ${\cal P}_{XY}$ is read off as
\begin{eqnarray}
\label{eq:nlsm_PS}%\eqref{eq:nlsm_PS}
{\cal P}_{XY}(k)&=&2{\cal P}_{\rm ini}^N\int \frac{d^3q}{(2\pi)^3}F_X(q,|\vec{k}-\vec{q}|)F^*_Y(q,|\vec{k}-\vec{q}|)\nonumber\\
&=&\frac{{\cal P}_{\rm ini}^N}{{2\pi^2}}\int dq\ q^2\int d\mu\ F_X(q,|\vec{k}-\vec{q}|)F^*_Y(q,|\vec{k}-\vec{q}|),
\end{eqnarray}
where $\mu=\hat{k}\cdot\hat{q}$.

\bibliography{MFFSOSF}

\end{document}